\documentclass[twocolumn]{aastex631}

\usepackage{graphicx}
\usepackage{bm}
\usepackage{hyperref}
\usepackage{mathtools}
\usepackage{array}
\usepackage{tabularx}
\usepackage{multirow}

\shorttitle{CCs, Pantheon+ SNe Ia, and QSOs favor coasting cosmologies}

\begin{document}

\title{Cosmic chronometers, Pantheon+ supernovae, and quasars \\
favor coasting cosmologies over the flat $\Lambda$CDM model}

\author[0000-0001-7576-0141]{Peter Raffai}
\affiliation{Institute of Physics and Astronomy, ELTE E\"otv\"os Lor\'and University, 1117 Budapest, Hungary}
\affiliation{HUN-REN–ELTE Extragalactic Astrophysics Research Group, 1117 Budapest, Hungary}

\author[0009-0003-2523-6474]{Adrienn Pataki}
\affiliation{Institute of Physics and Astronomy, ELTE E\"otv\"os Lor\'and University, 1117 Budapest, Hungary}

\author[0000-0001-7015-6551]{Rebeka L. B\"ottger}
\affiliation{Institute of Physics and Astronomy, ELTE E\"otv\"os Lor\'and University, 1117 Budapest, Hungary}

\author[0000-0002-6389-3542]{Alexandra Karsai}
\affiliation{Institute of Physics and Astronomy, ELTE E\"otv\"os Lor\'and University, 1117 Budapest, Hungary}

\author[0000-0003-3258-5763]{Gergely D\'alya}
\affiliation{L2IT, Laboratoire des 2 Infinis - Toulouse, Universit\'e de Toulouse, CNRS/IN2P3, UPS, F-31062 Toulouse Cedex 9, France}
\affiliation{Department of Physics and Astronomy, Universiteit Gent, B-9000 Ghent, Belgium}

\correspondingauthor{Peter Raffai}
\email{peter.raffai@ttk.elte.hu}

\begin{abstract}

We test and compare coasting cosmological models with curvature parameters ${k=\left\{ -1,0,+1 \right\}}$ in ${H_0^2 c^{-2}}$ units and the flat $\Lambda$CDM model by fitting them to cosmic chronometers (CC), the Pantheon+ sample of type Ia supernovae (SNe), and standardized quasars (QSOs). We used the \texttt{emcee} code for fitting CC data, a custom Markov Chain Monte Carlo implementation for SNe and QSOs, and Anderson-Darling tests for normality on normalized residuals for model comparison. Best-fit parameters are presented, constrained by data within redshift ranges $z\leq 2$ for CCs, $z\leq 2.3$ for SNe, and $z\leq 7.54$ for QSOs. Coasting models, particularly the flat coasting model, are generally favored over the flat $\Lambda$CDM model. The overfitting of the flat $\Lambda$CDM model to Pantheon+ SNe and the large intrinsic scatter in QSO data suggest a need to refine error estimates in these datasets. We also highlight the seemingly fine-tuned nature of either the CC data or $\Omega_{\mathrm{m},0}$ in the flat $\Lambda$CDM model to an ${H_1=H_0}$ coincidence when fitting ${H(z)=H_1z+H_0}$, a natural feature of coasting models.

\end{abstract}

\section{Introduction} 
\label{sec:intro}

Coasting cosmologies are a family of cosmological models in which the $a(t)$ scale factor grows linearly with cosmic time $t$ (for a review, see \citealt{Casado_2020}). This family includes {\it strictly linear} models where $a(t)\propto t$ at all times, and {\it quasi-linear} models
which propose an early universe evolution in accordance with the concordance model (i.e., Lambda Cold Dark Matter or $\Lambda$CDM, see \citealt{Peebles_Ratra_2003} for a review), with a transition to linear expansion sometime after recombination. Variations within coasting cosmologies arise from differences in the underlying assumptions driving the linear expansion and the value of the $k$ spatial curvature parameter they propose or allow. For instance, the earliest coasting model, developed by Arthur Milne in the 1930s~\citep{Milne_1935}, exhibits dynamics akin to an empty universe with zero $\Lambda$ cosmological constant and negative $k$. The $R_{\mathrm{h}}=ct$ model~\citep{Melia_2007,Melia_Shevchuk_2012,Melia_2020a} and the eternal coasting model by John and Joseph~\citep{John_Joseph_1996,John_Joseph_2000}, suggest $k=0$ and $k=+1$, respectively, although they permit any $k$ value (see, e.g., \citealt{John_Joseph_2000,John_Joseph_2023}). The hyperconical universe model introduced by Monjo~\citep{Monjo_2017,Monjo_2024} proposes linear expansion with $k=+1$ (for a comprehensive review on coasting models, see \citealt{Casado_2020}).

Despite the preeminent explanatory power of the $\Lambda$CDM model, tensions between the locally measured Hubble constant ($H_0$;~\citealt{Riess_2020}) and structure growth parameter ($S_8$;~\citealt{Di_Valentino_et_al_2021}) and those derived from cosmic microwave background (CMB) observations using the $\Lambda$CDM model~\citep{Planck_2018}, as well as other anomalies~\citep{Perivolaropoulos_Skara_2022}, suggests the necessity of exploring alternative cosmologies. Coasting models fit remarkably well to a broad range of cosmological datasets at low redshifts (see, e.g., Table 2 in \citealt{Melia_2018} and references therein). Strictly linear models offer natural solutions to several theoretical challenges in the $\Lambda$CDM model, including the horizon, flatness, cosmological constant, synchronicity, cosmic coincidence, and cosmic age problems (see \citealt{Casado_2020} for a review). The horizon and flatness problems are addressed within the $\Lambda$CDM framework through the theory of cosmic inflation~\citep{Guth_1981,Baumann_2009}, and other issues listed may represent unlikely coincidences in the realizations of $\Lambda$CDM model parameters. Strictly linear models, however, face significant challenges in explaining observed phenomena presumably set by pre-recombination physics, such as light element abundances inherited from primordial nucleosynthesis~\citep{Kaplinghat_et_al_1999,Sethi_et_al_1999,Kaplinghat_et_al_2000,Lewis_et_al_2016} and CMB anisotropies (see, e.g., \citealt{Fujii_2020,Melia_2020b,Melia_2022}), both of which are well explained by the $\Lambda$CDM framework~\citep{Dodelson_2003}. Quasi-linear coasting models (e.g.,~\citealt{Kolb_1989}) are identical to the $\Lambda$CDM model before recombination, and thus are free from these challenges.

In~\citet{Raffai_et_al_2024} we used gravitational-wave standard sirens~\citep{Holz_Hughes_2005} observed in the first three observing runs~\citep{Abbott_et_al_2023a} of the LIGO-Virgo-KAGRA detector network~\citep{Aasi_et_al_2015,Acernese_et_al_2015,Akutsu_et_al_2021} to constrain $H_0$ for coasting cosmologies with three fixed values of ${k=\left\{ -1,0,+1 \right\}}$ in ${H_0^2 c^{-2}}$ units, where $c$ is the speed of light in vacuum. From a combined analysis of $46$ dark sirens and a single bright siren~\citep{Abbott_et_al_2023b}, we obtained maximum posteriors and $68.3\%$ highest density intervals of ${H_0=\left\{68.1^{+8.5}_{-5.6},67.5^{+8.3}_{-5.2},67.1^{+6.6}_{-5.8} \right\}~\mathrm{km\ s^{-1}\ Mpc^{-1}}}$, respectively. Our results constrained coasting models in the redshift range of $z\lesssim 0.8$, concluding that, within the broad ranges of measurement and source population modeling uncertainties, the tested coasting models and the flat ($k=0$) $\Lambda$CDM model fit equally well to the gravitational-wave detections.

In this paper, we test and constrain the coasting models with ${k=\left\{ -1,0,+1 \right\}}$ in ${H_0^2 c^{-2}}$ units and the flat $\Lambda$CDM model using cosmic chronometers (CCs), type Ia supernovae (SNe Ia), and quasars (QSOs). In~\citet{Raffai_et_al_2024} we introduced ${H_0=62.41^{+2.95}_{-2.96}~\mathrm{km~s^{-1}~Mpc^{-1}}}$ as a reference for coasting models regardless of their $k$, determined from CCs with the differential age method~\citep{Jimenez_Loeb_2002,Simon_Verde_Jimenez_2005}. In Section~\ref{sec:Cosmic_chronometers}, we describe the details of how we obtained this reference $H_0$. In Section~\ref{sec:SNIa} and Section~\ref{sec:QSO}, we present test results for SNe and QSOs. We discuss our findings and draw conclusions in Section~\ref{sec:Conclusion}.

\section{Tests with cosmic chronometers}
\label{sec:Cosmic_chronometers}

In most cosmologies, including the $\Lambda$CDM and coasting models, $a(t)$ can be expressed as a function of the cosmological redshift $z$ of photons emitted by sources at cosmic time $t$ as ${a(z)=(1+z)^{-1}}$. In coasting cosmologies, ${a(t)=H_0t}$, and thus ${H(t)\equiv \dot{a}a^{-1}=t^{-1}}$ (where the overdot indicates a derivative with respect to $t$) and ${H(z)=H_0a^{-1}=H_0(1+z)}$ in the $t$ and $z$ range where linear expansion is assumed. In contrast, 
\begin{equation}\label{Eq:Hz_LCDM}
H(z)=H_0 \sqrt{\Omega_\mathrm{m,0}\left( 1+z \right)^3+1-\Omega_\mathrm{m,0}}
\end{equation}
in the flat $\Lambda$CDM model, where $\Omega_\mathrm{m,0}$ is the matter density parameter today, and the contribution of radiation to the total density is neglected. We can also derive ${H(z)=-\dot{z}\left( 1+z \right)^{-1}}$ for all cosmologies that satisfy ${a(z)=(1+z)^{-1}}$. The differential age method~\citep{Jimenez_Loeb_2002,Simon_Verde_Jimenez_2005} takes advantage of the fact that $\dot{z}$ at $z$ can in practice be approximated as ${\dot{z}\approx \Delta z \Delta t^{-1}}$, where $\Delta z$ and $\Delta t$ are the redshift and age differences of, e.g., pairs of galaxies with redshifts around $z$. Passively evolving galaxies allow measuring their $\Delta t$ age differences from observed differences in their stellar populations, from which $H(z)$ can be determined with uncertainties typically dominated by uncertainties of the $\Delta t$ measurement. Such objects utilized in measuring $H(z)$ are usually referred to as \emph{cosmic chronometers} (CCs).

\begin{figure}
 	\includegraphics[width=\columnwidth]{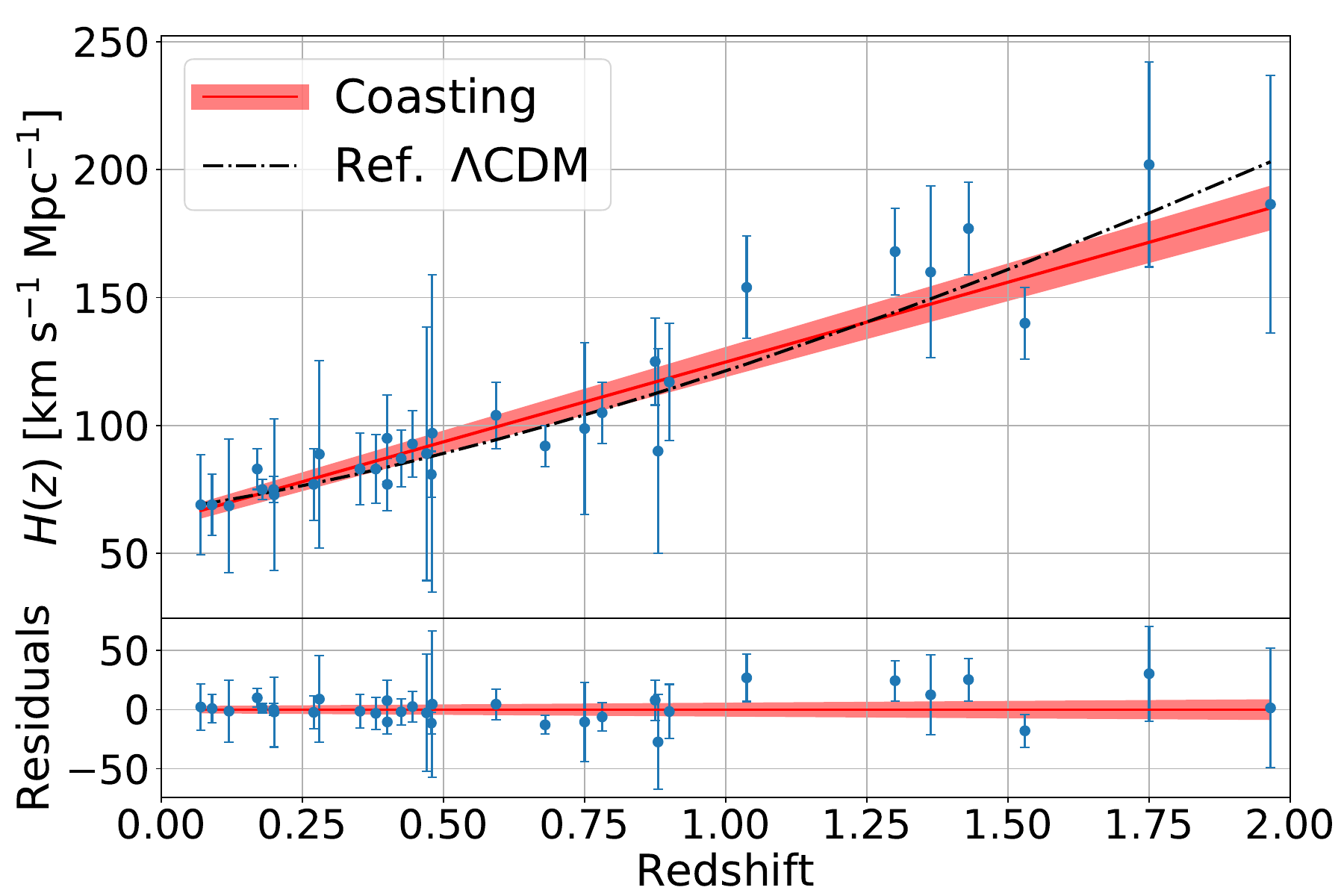}
     \caption{The $32$ CC data points from Table 1 of \citet{Moresco_et_al_2022} (blue dots with error bars), along with the best-fit coasting $H(z)$ (red solid) line plotted with $1\sigma$ confidence bands. For comparison, we also plotted the $\Lambda$CDM $H(z)$ curve with the Planck+BAO cosmological parameters from \citet{Planck_2018} (upper panel, black dash-dotted curve). The lower panel shows the residuals after subtracting the best-fit coasting $H(z)$ line from the measured data points. Parameters of the coasting and reference $\Lambda$CDM models are given in Table~\ref{tab:table1}.}
     \label{fig:fig1}
\end{figure}

\citet{Melia_Maier_2013} determined $H_0$ for coasting models with CCs by fitting ${H(z)=H_0(1+z)}$ to $19$ $H(z)$ measurements from \citet{Simon_et_al_2005,Stern_et_al_2010,Moresco_et_al_2012}, and obtained ${H_0=63.2\pm 1.6~\mathrm{km~s^{-1}~Mpc^{-1}}}$ regardless of $k$. We updated their result by fitting the $H(z)$ formula in coasting models to the $32$ $H(z)$ measurements~\citep{Simon_et_al_2005,Stern_et_al_2010,Moresco_et_al_2012,Zhang_et_al_2014,Moresco_2015,Moresco_et_al_2016,Ratsimbazafy_et_al_2017,Borghi_et_al_2022} summarized in Table 1 of \citet{Moresco_et_al_2022} using the \texttt{emcee}\footnote{\label{note:emcee_code}\url{https://gitlab.com/mmoresco/CCcovariance}}~\citep{Foreman_Mackey_et_al_2013} Markov Chain Monte Carlo (MCMC) code with the full statistical and systematic covariance matrix of the data. For comparisons, we also determined the best-fit parameters for a flat $\Lambda$CDM and a two-parameter ${H(z)=H_1z+H_0}$ model using the same data and code. We show the $32$ CC data points with $1\sigma$ error bars in Figure~\ref{fig:fig1}, along with the best-fit coasting $H(z)$ line and a reference $\Lambda$CDM $H(z)$ curve we obtained with the Planck+BAO cosmological parameters from \citet{Planck_2018}. We present the best-fit parameters for the tested models in Table~\ref{tab:table1} as the medians of the parameter posteriors, with averaged $1\sigma$ errors.

\begin{table}[b]
\caption{\label{tab:table1} Model Fit and Test Results for CC Data}
\begin{ruledtabular}
\begin{tabular}{lccc}
\textrm{Fitted model}&
\textrm{Best-fit parameters}&
\textrm{$\chi^2$}&
\textrm{$\log_{10}p^{-1}$}\\
\colrule
Coasting & $H_0=62.41\pm 2.96$ & 16.73 & $0.733^*$\\
$H_1z+H_0$ & $H_0=62.51\pm 5.29$ & 16.73 &  \\
 & $H_1=62.28\pm 6.02$ & & \\
Flat $\Lambda$CDM & $H_0=66.70\pm 5.41$ & 14.57 & $1.062^*$\\
 & $\Omega_{\mathrm{m},0} = 0.33\pm 0.07$ &  & \\
Ref. $\Lambda$CDM & $H_0=67.66$ (fixed) & 14.62 &   \\
 & $\Omega_{\mathrm{m},0}=0.311$ (fixed) &  & \\
 \end{tabular}
\tablecomments{$H_{\{0,1\}}$ values are given in $[\mathrm{km}\ \mathrm{s}^{-1}\ \mathrm{Mpc}^{-1}]$ units. The fixed parameters of the reference $\Lambda$CDM model were taken from \citet{Planck_2018}. The $\chi^2$ values were calculated with the \texttt{emcee} code~\citep{Foreman_Mackey_et_al_2013} we used for model fitting. The $p$-values are results of the AD test we used for model testing. Asterisks mark models that are consistent with the null hypothesis of being the true model (i.e., $p\geq 0.05$ or $\log_{10}p^{-1}\leq 1.301$). We left the ${\log_{10}p^{-1}}$ column blank for the $H_1z+H_0$ and Ref. $\Lambda$CDM models because we did not include them in the model comparison. Plots of posterior distributions are available in our public code repository~\citep{Zenodo_repo}.}
\end{ruledtabular}
\end{table}

The fitted models are linear in $H_0$, but the flat $\Lambda$CDM model is nonlinear in $\Omega_{\mathrm{m},0}$. Applying priors or external calibrators in fitting often causes models to become nonlinear in their fit parameters~\citep{Andrae_et_al_2010}. For models that are nonlinear in one or more fit parameters, the numbers of degrees of freedom (and thus the reduced chi-squares) cannot be estimated reliably. Additionally, the uncertainty in $\chi^2$ limits its applicability in model comparisons, especially when the number of data points is low~\citep{Andrae_et_al_2010}. Because of these considerations, following the recommendations in \citet{Andrae_et_al_2010}, we applied an Anderson-Darling (AD) test~\citep{Anderson_Darling_1952,adtest_2024} to check if our normalized fit residuals follow a standard normal distribution, which should be the case if a model is the true model underlying the data, and the data, after the necessary cleaning, preprocessing and error estimation procedure, is no longer contaminated with non-cosmological effects. We used the AD test $p$-values for model testing and ${\log_{10}\mathcal{B}=\log_{10}(p_\mathrm{max}/p)}$ Bayes factors for model comparisons, with $p_\mathrm{max}$ as the highest $p$-value across all models, while retaining $\chi^2$-minimization for model fitting. We show the base 10 logs of the reciprocals of AD test $p$-values for the tested models in Table~\ref{tab:table1}.

We have found no data points deviating by more than $3\sigma$ from the best-fit models or from the reference $\Lambda$CDM model. Models for which the AD test $p$-value is ${p\geq 0.05}$ (${\log_{10}p^{-1}\leq 1.301}$) are considered consistent with the null hypothesis of being the true model. We found that all three coasting models and the flat $\Lambda$CDM model satisfy this condition, with the coasting models being slightly favored by the CC data (${\log_{10}\mathcal{B}=0.329}$), although the preference is not significant.

As Table~\ref{tab:table1} shows, the best-fit parameters for the flat $\Lambda$CDM model are consistent within $1\sigma$ with parameters of the reference $\Lambda$CDM model. For the ${H(z)=H_1z+H_0}$ model, the best-fit $H_0$ and $H_1$ are equal to each other within $\sim 0.4\%$ and they match the best-fit $H_0$ for coasting models within $\sim 0.2\%$ (with $\chi^2$ values matching to the given number of decimals). These results are curious given that (i) a flat $\Lambda$CDM universe does not imply the consistency of ${H(z)=H_1z+H_0}$ with data extending to $z\gtrsim 1$, and (ii) even for $z\ll 1$, an effective $H_1$ in a flat $\Lambda$CDM universe depends on the unrelated values of $H_0$ and $\Omega_{\mathrm{m},0}$ (${H_1=3H_0\Omega_{\mathrm{m},0}/2}$), which does not imply that the best-fit $H_1$ and $H_0$ should match. In contrast, these coincidences would be natural if the true cosmological model was a coasting one (at least in the $z\lesssim 2$ redshift range of the fitted CC data).

To evaluate the significance of the ${H_1=H_0}$ coincidence, we kept the $z$ redshifts and the $H(z)$ errors of the $32$ actual CC measurements and simulated two sets of one million mock samples of $32$ $H(z)$ measurements in $\Lambda$CDM universes described by Equation~(\ref{Eq:Hz_LCDM}), with $H_0$ and $\Omega_{\mathrm{m},0}$ values randomized from normal distributions. The normal distributions in the first set were defined by the flat $\Lambda$CDM best-fit parameters in Table~\ref{tab:table1}, and in the second set by the reference $\Lambda$CDM parameters ${H_0=67.66\pm 0.42~\mathrm{km~s^{-1}~Mpc^{-1}}}$ and ${\Omega_{\mathrm{m},0}=0.3111\pm 0.0056}$ from \citet{Planck_2018}. We fitted the $H_1z+H_0$ model to the two sets and found that only $2\%$ and $10\%$ of the one million $(H_1,H_0)$ pairs obtained from the fits were equal within $0.4\%$ for the flat and reference $\Lambda$CDM models, respectively. We then created mock samples of $32$ $H(z)$ measurements again with the $z$ and $H(z)$ errors of the real CC data, but calculated with fixed $H_0$ in Equation~(\ref{Eq:Hz_LCDM}) and varied $\Omega_{\mathrm{m},0}$ over the interval ${\Omega_{\mathrm{m},0}\in [0,1.5]}$. We found that regardless of what $H_0$ value we choose in Equation~(\ref{Eq:Hz_LCDM}), only in the narrow interval of ${\Omega_{\mathrm{m},0}\in [0.317,0.320]}$ do we get ${H_1=H_0}$ coincidences within $0.4\%$ precision, with the closest match at ${\Omega_{\mathrm{m},0}=0.3186}$. Note that $\Omega_{\mathrm{m},0}$ in the reference $\Lambda$CDM model is consistent within $\sim 1\sigma$ with the narrow $\Omega_{\mathrm{m},0}$ interval corresponding to the ${H_1=H_0}$ coincidence. This seemingly {\it fine-tuned} nature of either the CC data or $\Omega_{\mathrm{m},0}$ in the $\Lambda$CDM model to an ${H_1=H_0}$ coincidence adds to the arguments for studying coasting models and makes our findings worth following up with an improved CC dataset in the future.

\section{Tests with type Ia supernovae}
\label{sec:SNIa}

We further tested and constrained the three coasting models and the flat $\Lambda$CDM model using the Pantheon+ sample of SNe Ia~\citep{Scolnic_et_al_2022}, compiled from $1701$ light curves of $1550$ SNe within $z\lesssim 2.3$. \citet{Scolnic_et_al_2022} fitted the light curves with the SALT2 model~\citep{Guy_et_al_2007} following \citet{Brout_et_al_2022a}. We used the $x_0$ light-curve amplitudes as ${m_B\equiv -2.5\log_{10}(x_0)}$, the $x_1$ stretch and the $c$ color parameters from their fit data\footnote{\label{note:Pantheon_data}See Pantheon+ data release at \url{https://github.com/PantheonPlusSH0ES/DataRelease}.}. Following \citet{Brout_et_al_2022b}, who built on \citet{Tripp_1998} and \citet{Kessler_Scolnic_2017}, we inferred the distance moduli of the SNe as:
\begin{equation}\label{eq:dist_mod}
\mu_\mathrm{SN}=m_B+\alpha x_1-\beta c - M_B -\delta_\mathrm{bias} + \delta_\mathrm{host}
\end{equation}
where $\alpha$ and $\beta$ are global nuisance parameters, $M_B$ is the fiducial magnitude of an SN, and $\delta_\mathrm{bias}$ is a correction term accounting for selection biases\footref{note:Pantheon_data} (see \citealt{Brout_et_al_2022b} and \citealt{Popovic_et_al_2021} for details). ${\delta_\mathrm{host}\equiv \gamma \tilde{\delta}_\mathrm{host}(M_\star)}$ corrects for residual correlations between the standardized brightness of an SN and the host-galaxy stellar mass ($M_\star$), $\gamma$ is a nuisance parameter, and $\tilde{\delta}_\mathrm{host}$ is a function of $M_\star$ defined in Equation (2) of \citet{Brout_et_al_2022b} and \citet{Popovic_et_al_2021}. The $M_\star$ values for the SN hosts\footref{note:Pantheon_data} are presented in \citet{Scolnic_et_al_2022} and references therein. The distance modulus of a source at redshift $z$ is
\begin{equation}\label{eq:dist_mod_model}
\mu(z)\equiv 5\log_{10}\left( \frac{d_L(z)}{10\ \mathrm{pc}} \right)
\end{equation}
where $d_L$ is the luminosity distance of the source. For the three coasting cosmologies, $d_L$ is 
\begin{equation}\label{Eq:dL}
d_L(z)=\frac{c}{H_0}\left( 1+z \right)
\begin{cases} 
\sinh(\ln(1+z)) &\textrm{for $k=-1$}\\
\ln(1+z) &\textrm{for $k=0$}\\
\left| \sin(\ln(1+z)) \right| &\textrm{for $k=+1$}\\
\end{cases}
\end{equation}
and for the flat $\Lambda$CDM model, with $H(z)$ given in Equation~(\ref{Eq:Hz_LCDM}), it is
\begin{equation}\label{eq:dL_LCDM}
d_L(z)=c\left( 1+z \right)\int_{0}^{z}\frac{\mathrm{d}z'}{H(z')}.
\end{equation}

We fitted the nuisance parameters $\alpha$, $\beta$, $\gamma$, $M_B$ and the cosmological parameters $H_0$ and $\Omega_\mathrm{m,0}$ simultaneously using our own MCMC code~\citep{Zenodo_repo} minimizing
\begin{equation}\label{eq:chi_SNe}
\chi^2=\Delta \bm{D}^{T}C^{-1}_\mathrm{stat+syst} \Delta \bm{D}
\end{equation}
where $\Delta \bm{D}$ is the vector of $1701$ SN Ia distance-modulus residuals computed as
\begin{equation}\label{eq:Di}
\Delta D_i=\mu_{\mathrm{SN},i}-\mu (z_i).
\end{equation}
$\mu_{\mathrm{SN},i}$ and $\mu (z_i)$ are the distance moduli for the $i$th SN calculated using Eqs.~(\ref{eq:dist_mod}) and~(\ref{eq:dist_mod_model}), respectively, and $C^{-1}_\mathrm{stat+syst}$ is the inverse of the covariance matrix accounting for both statistical and systematic uncertainties\footref{note:Pantheon_data} (see \citealt{Brout_et_al_2022b}). For $z_i$, we used the cosmological redshift of the SN host galaxy in the CMB frame, corrected for peculiar velocity (denoted as $z_\mathrm{HD}$ in Pantheon+\footref{note:Pantheon_data}; see \citealt{Carr_et_al_2022}).

We can see from Eqs.~(\ref{eq:dist_mod}) and~(\ref{eq:dist_mod_model}) that $M_B$ and $H_0$ are degenerate and cannot be fitted independently. To address this, we followed \citet{Brout_et_al_2022b} by using the Cepheid-calibrated host-galaxy distance moduli $\mu^\mathrm{Cepheid}_i$ from SH0ES~\citep{Riess_et_al_2022} as external calibrators, replacing $\mu (z_i)$ with $\mu^\mathrm{Cepheid}_i$ in Eq.~(\ref{eq:Di}) for the $77$ SNe where available. We used the $C_\mathrm{stat+syst}$ in Eq.~(\ref{eq:chi_SNe}) that included the Cepheid host-distance covariance matrix from \citet{Riess_et_al_2022}.

Similarly to \citet{Amanullah_et_al_2010}, \citet{Riess_et_al_2022}, and others, we applied the method of {\it sigma clipping} in model fitting, i.e., we iteratively removed data points that deviated from the global fits by more than $3\sigma$ until no such outliers remained. As shown by, e.g., \citet{Kowalski_et_al_2008} with simulations, this technique preserves the fit results in the absence of contamination and reduces the impact of any contaminating data. Out of the $1701$ SN data points, sigma clipping removed ${N=\left\{ 17,18,24 \right\}}$ for the ${k=\left\{ -1,0,+1 \right\}}$ coasting models and $N=15$ for the flat $\Lambda$CDM model. 

We present the best-fit parameters as the medians of the posteriors, along with their 16th and 84th percentile errors, in Table~\ref{tab:table2}. We give the $\chi^2$ values in Table~\ref{tab:table2} calculated for all $1701$ SN observations. Our best-fit $H_0$ and $\Omega_{\mathrm{m},0}$ for the flat $\Lambda$CDM model are consistent with ${H_0=73.6\pm 1.1~\mathrm{km~s^{-1}~Mpc^{-1}}}$ (within $0.3\sigma$) and ${\Omega_{\mathrm{m},0}=0.334\pm 0.018}$ (within $0.2\sigma$) in \citet{Brout_et_al_2022b}. Our $H_0$ is also consistent with ${H_0=73.30\pm 1.04~\mathrm{km~s^{-1}~Mpc^{-1}}}$ (within $0.1\sigma$) and ${H_0=73.04\pm 1.04~\mathrm{km~s^{-1}~Mpc^{-1}}}$ (within $0.06\sigma$) in \citet{Riess_et_al_2022}, obtained with and without including high-redshift ($z\in[0.15,0.8)$) SNe in their analysis, respectively.

\begin{table}
\caption{\label{tab:table2} Model Fit and Test Results for SN Ia Data}
\begin{ruledtabular}
\begin{tabular}{lccc}
\multicolumn{4}{c}{\textrm{Coasting}} \\
\colrule
 & \textrm{Flat} & \textrm{Closed} & \textrm{Open} \\
\colrule
$\alpha$ & $0.144_{-0.004}^{+0.004}$ & $0.144_{-0.004}^{+0.004}$ & $0.146_{-0.004}^{+0.004}$ \\
$\beta$ & $2.907_{-0.072}^{+0.072}$ & $2.887_{-0.073}^{+0.072}$ & $2.949_{-0.072}^{+0.072}$ \\
$\gamma$ & $0.008_{-0.011}^{+0.011}$ & $0.004_{-0.011}^{+0.011}$ & $0.011_{-0.011}^{+0.011}$ \\
$M_B$ & $-19.187_{-0.031}^{+0.031}$ & $-19.177_{-0.030}^{+0.031}$ & $-19.198_{-0.031}^{+0.031}$ \\
$H_0$ & $71.39_{-1.01}^{+1.02}$ & $71.02_{-0.99}^{+1.01}$ & $71.71_{-1.02}^{+1.04}$ \\
\colrule
$\chi^2$ & \textrm{1820} & \textrm{1981} & \textrm{1775} \\
$\log_{10}\mathcal{B}$ & \textrm{0.003} & \textrm{0 (1.950)} & \textrm{1.207} \\
\toprule
\multicolumn{4}{c}{\textrm{Flat $\Lambda$CDM}} \\
\colrule
& $\alpha$ & $0.149_{-0.004}^{+0.004}$ & \\
& $\beta$ & $2.979_{-0.072}^{+0.070}$ & \\
& $\gamma$ & $0.010_{-0.011}^{+0.011}$ & \\
& $M_B$ & $-19.210_{-0.031}^{+0.031}$ & \\
& $H_0$ & $73.13_{-1.05}^{+1.08}$ & \\
& $\Omega_\mathrm{m,0}$ & $0.328_{-0.018}^{+0.018}$ & \\
\colrule
& $\chi^2$ & \textrm{1755} & \\
& $\log_{10}\mathcal{B}$ & \textrm{2.502} & \\
\end{tabular}
\tablecomments{$H_0$ values are given in $[\mathrm{km}\ \mathrm{s}^{-1}\ \mathrm{Mpc}^{-1}]$ units. ${\log_{10}p^{-1}_\mathrm{max}=1.950}$ (for the closed coasting model) is shown in brackets in the ${\log_{10}\mathcal{B}=\log_{10}(p_\mathrm{max}/p)}$ row. Plots of posterior distributions are available in our public code repository~\citep{Zenodo_repo}.}
\end{ruledtabular}
\end{table}

Our best-fit ${H_0=73.13^{+1.08}_{-1.05}~\mathrm{km~s^{-1}~Mpc^{-1}}}$ for the flat $\Lambda$CDM model is in $4.8\sigma$ tension with the Planck+BAO ${H_0=67.66\pm 0.42~\mathrm{km~s^{-1}~Mpc^{-1}}}$ \citep{Planck_2018}. Although the best-fit $H_0$ values for the ${k=\left\{ -1,0,+1 \right\}}$ coasting models are ${\left\{3.7\sigma,3.4\sigma,3.1\sigma \right\}}$ above the Planck+BAO $H_0$, we cannot claim that these models necessarily alleviate the tension between the CMB- and SN-based $H_0$ measurements. Strictly linear coasting models face significant challenges in explaining CMB anisotropies, preventing their fit to CMB data. However, quasi-linear versions of the ${k=\left\{ -1,0,+1 \right\}}$ coasting models, with linear expansion in the $z\leq 2.3$ redshift range of Pantheon+ SNe and $\Lambda$CDM-type expansion at earlier times, including the time of recombination, may yield a CMB-based $H_0$ value similar to that of the $\Lambda$CDM model, potentially alleviating the $H_0$ tension.

\begin{figure}
 	\includegraphics[width=\columnwidth]{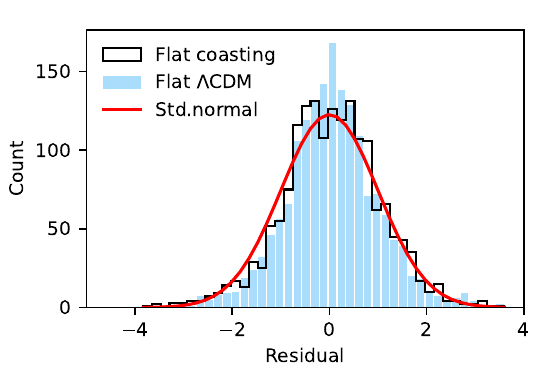}
        \includegraphics[width=\columnwidth]{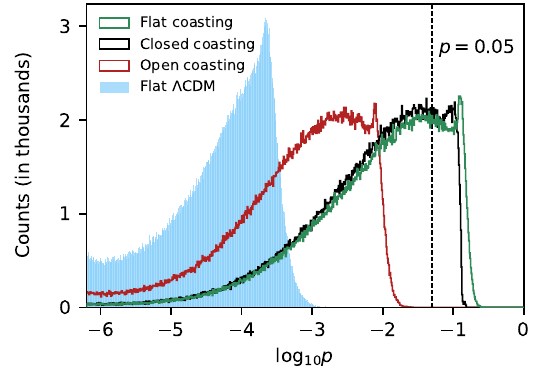}
     \caption{Upper panel: Histograms of normalized residuals for the best-fit flat coasting and flat $\Lambda$CDM models, after fitting the Pantheon+ SN Ia data~\citep{Scolnic_et_al_2022}. For both models, we excluded five outliers exceeding $4\sigma$ from the $1701$ data points. The red curve shows the standard normal distribution expected for the true model underlying the data. Low-value residuals are clearly overrepresented in the $\Lambda$CDM sample, with the sample standard deviation being $\sigma=0.95$ compared to $\sigma=1.00$ for the flat coasting sample. Lower panel: Histograms of AD-test ${\log_{10}p}$ values for half a million realizations of flat $\Lambda$CDM and coasting models, sampled from the joint posterior distributions obtained from fitting to $1701$ Pantheon+ SN Ia observations. Realizations with $p\geq 0.05$ (${\log_{10}p=-1.301}$, dashed line) are consistent with the null hypothesis of being the true model: $0\%$ for flat $\Lambda$CDM, and ${\left\{ 0\%,19\%,23\% \right\}}$ for open, closed, and flat coasting models, respectively.}
     \label{fig:fig2}
\end{figure}

We applied the same AD test for model comparison as in Section~\ref{sec:Cosmic_chronometers}. Since practically none of the $1701$ SN data points should deviate from a cosmological model by more than $4\sigma$, we excluded such outliers from the AD-tested samples, assuming non-cosmological contamination. The five excluded data points were the same for all tested models. None of the best-fit models met the $p\geq 0.05$ criterion for consistency with the SN data, with the closest ones being $p=0.01$ for the closed and flat coasting models (${\log_{10}p^{-1}_\mathrm{max}=1.950}$ and ${\log_{10}p^{-1}=1.953}$, respectively; see Table~\ref{tab:table2}). The closed and flat coasting models are strongly favored over the open coasting (${\log_{10}\mathcal{B}\simeq 1.2}$) and the flat $\Lambda$CDM models (${\log_{10}\mathcal{B}\simeq 2.5}$). In light of the lower $\chi^2$ values for the flat $\Lambda$CDM ($\chi^2=1755$) and open coasting ($\chi^2=1775$) models than for the flat and closed coasting models ($\chi^2=1820$ and $\chi^2=1981$, respectively; see Table~\ref{tab:table2}), these preferences suggest that the former two models overfit the SN data. The upper panel of Figure~\ref{fig:fig2} shows the histograms of normalized residuals for the best-fit flat coasting and flat $\Lambda$CDM models, with a reference curve for the standard normal distribution expected for the true model underlying the SN data. Low-value residuals are indeed overrepresented in the $\Lambda$CDM sample, with the sample standard deviation being $\sigma=0.95$ compared to $\sigma=1.00$ for the flat coasting sample. In the lower panel of Figure 2, we present histograms of AD-test ${\log_{10}p}$ values for half a million realizations of flat $\Lambda$CDM and coasting models, sampled from the joint posterior distributions fitted to the $1701$ Pantheon+ SN Ia observations. None of the flat $\Lambda$CDM realizations satisfy the $p\geq 0.05$ condition for consistency with the Pantheon+ sample, while ${\left\{ 0\%,19\%,23\% \right\}}$ of the open, closed, and flat coasting model realizations do. Additionally, $p$-values exceed the highest (median) $p$-value of the flat $\Lambda$CDM realizations for ${\left\{ 25\%,70\%,71\% \right\}}$ (${\left\{ 84\%,96\%,96\% \right\}}$) of open, closed, and flat coasting model realizations, respectively. These results show a clear overall preference for coasting models over the flat $\Lambda$CDM model by the Pantheon+ SNIa sample.

The flat $\Lambda$CDM model being less favored by the SN data is not due to sigma clipping. Without the iterative outlier rejection, the $k=+1$ coasting model is the most favored by data, with $p=0.014$. The flat coasting model is only slightly less favored (${\log_{10}\mathcal{B}\simeq 0.2}$), while the strong preference over the flat $\Lambda$CDM (${\log_{10}\mathcal{B}\simeq 2.8}$) and open coasting models (${\log_{10}\mathcal{B}\simeq 1.2}$) remains. Without sigma clipping, no flat $\Lambda$CDM realizations satisfy $p\geq 0.05$, while ${\left\{ 0\%,21\%,19\% \right\}}$ of open, closed, and flat coasting model realizations do. The highest (median) $p$-value for flat $\Lambda$CDM is exceeded by ${\left\{ 42\%,81\%,75\% \right\}}$ (${\left\{ 87\%,98\%,96\% \right\}}$) of $p$-values for open, closed, and flat coasting models, respectively.

To provide broader context for Table~\ref{tab:table2}, we also fitted a coasting model to SN data for which we allowed $k$ to vary freely between $-2$ and $2$ in ${H_0^2 c^{-2}}$ units. The best-fit $k$ is ${k=-1.044^{+0.139}_{-0.140}}$, with other parameters and $\chi^2$ differing by no more than $0.06\%$ from those in Table~\ref{tab:table2} for the open ($k=-1$) coasting model. The model yields ${\log_{10}\mathcal{B}=1.240}$, also very close to the ${\log_{10}\mathcal{B}=1.207}$ of the open coasting model. We have made the posterior distribution plots available in our public code repository~\citep{Zenodo_repo}.

\section{Tests with quasars}
\label{sec:QSO}

The Pantheon+ compilation allows testing cosmologies only up to $z\simeq 2.3$. As \citet{Risaliti_Lusso_2015} point out, QSOs can serve as standardizable candles to extend this test to higher redshifts, even as high as $z\simeq 7.54$~\citep{Lusso_et_al_2020}. QSO standardization is based on the empirical relation $\log (L_\mathrm{X}) \propto \log (L_\mathrm{UV})$ between their rest-frame monochromatic luminosities at $2\ \mathrm{keV}$ and $2500\ \mathrm{\AA}$~\citep{Risaliti_Lusso_2015}. This relation allows the QSO distance modulus to be given as
\begin{equation}\label{eq:dist_mod_QSOs}
\mu_\mathrm{QSO}=\frac{5}{2(\gamma-1)}\left[ \log_{10}(F_\mathrm{X})-\gamma \log_{10}(F_\mathrm{UV})\right]-\beta
\end{equation}
where $F_\mathrm{X}$ and $F_\mathrm{UV}$ are the measured rest-frame fluxes at $2\ \mathrm{keV}$ and $2500\ \mathrm{\AA}$, respectively, and $\gamma$ and $\beta$ are global nuisance parameters~\citep{Risaliti_Lusso_2015}. These parameters can be calibrated using QSOs with known $z$, $F_\mathrm{X}$, and $F_\mathrm{UV}$ values against SNe Ia with known distance moduli (see, e.g.,~\citealt{Risaliti_Lusso_2015}), or by fitting them alongside the parameters of a cosmological model. We used the latter approach, fitting the coasting and flat $\Lambda$CDM models to QSO data from \citet{Lusso_et_al_2020} with the same MCMC code~\citep{Zenodo_repo} as in Section~\ref{sec:SNIa}. Since a full covariance matrix was not published with the QSO data, we minimized:
\begin{equation}\label{eq:Chi_square_QSO}
\chi^2=\sum_{i=1}^{\mathrm{QSOs}}\frac{\left[ \mu_{\mathrm{QSO},i}-\mu(z_i) \right]^2}{\sigma_{\mu,i}^{2}}
\end{equation}
where $\mu_{\mathrm{QSO},i}$ and $\mu (z_i)$ are the distance moduli for the $i$th QSO calculated using Eqs.~(\ref{eq:dist_mod_QSOs}) and~(\ref{eq:dist_mod_model}). The variance of $\mu_{\mathrm{QSO},i}$ is:
\begin{equation}\label{eq:sigma_QSO}
\sigma_{\mu,i}^2=\frac{25}{4}\frac{\sigma_{\mathrm{X},i}\sigma_{\mathrm{U},i}}{(\gamma-1)^2}\left( \frac{\sigma_{\mathrm{X},i}}{\sigma_{\mathrm{U},i}}+\gamma^2\frac{\sigma_{\mathrm{U},i}}{\sigma_{\mathrm{X},i}}-2\gamma r_{\{\mathrm{X},\mathrm{U}\}} \right)
\end{equation}
where $\sigma_{\mathrm{X},i}$ and $\sigma_{\mathrm{U},i}$ are the errors on $\log_{10}(F_\mathrm{X})$ and $\log_{10}(F_\mathrm{UV})$, respectively, and ${r_{\{\mathrm{X},\mathrm{U}\}}=0.776}$ is their sample Pearson correlation coefficient (note that \citealt{Lusso_et_al_2020} ignored the $r_{\{\mathrm{X},\mathrm{U}\}}$ term when fitting the same QSO data). We used the symmetrized $H_0$ posteriors from the SN fits (see Table~\ref{tab:table2}) as priors for the QSO fits. Due to degeneracy between $\beta$ and $H_0$ (cf. Equation~(\ref{eq:Chi_square_QSO}) with Eqs.~(\ref{eq:dist_mod_QSOs}) and~(\ref{eq:dist_mod_model})), their QSO posteriors are determined by these $H_0$ priors.

As in Section~\ref{sec:SNIa}, we applied iterative $>3\sigma$ outlier rejection (sigma clipping), also used by \citet{Bargiacchi_et_al_2021} in fitting the same QSO dataset. For the flat $\Lambda$CDM model, we fixed $\Omega_\mathrm{m,0}$ at its SN best-fit value, as the fit did not converge within the physical range ${\Omega_\mathrm{m,0}\in[0,1]}$ when $\Omega_\mathrm{m,0}$ was left free. Out of the $2421$ $\mu_\mathrm{QSO}$ data points, sigma clipping removed ${N=\left\{ 1573,1607,1649 \right\}}$ for the ${k=\left\{ -1,0,+1 \right\}}$ coasting models and $N=1606$ for flat $\Lambda$CDM, due to large intrinsic scatter. Note that including the covariance term in Equation~(\ref{eq:sigma_QSO}) reduces the $\sigma_{\mu,i}$ errors compared to when it is ignored (e.g., in \citealt{Lusso_et_al_2020}), resulting in more outliers being rejected by sigma clipping. Table~\ref{tab:table3} shows the best-fit parameters for the remaining $\sim 1/3$ of the data, presented as the medians of the posteriors with $16$th and $84$th percentile errors. The $\chi^2$ values in Table~\ref{tab:table3} were calculated using Equation~(\ref{eq:Chi_square_QSO}) and all $2421$ data points.

\begin{table}
\caption{\label{tab:table3} Model Fit Results for QSO Data}
\begin{ruledtabular}
\begin{tabular}{lccc}
\multicolumn{4}{c}{\textrm{Coasting (SNe $H_0$ posteriors used as $H_0$ priors)}} \\
\colrule
 & \textrm{Flat} & \textrm{Closed} & \textrm{Open} \\
\colrule
$\gamma$ & $0.707_{-0.002}^{+0.002}$ & $0.683_{-0.002}^{+0.002}$ & $0.718_{-0.002}^{+0.002}$ \\
$\beta$ & $56.675_{-0.224}^{+0.222}$ & $54.475_{-0.168}^{+0.195}$ & $57.789_{-0.208}^{+0.209}$ \\
$H_0$ & $71.38_{-1.01}^{+1.01}$ & $71.02_{-0.99}^{+0.99}$ & $71.72_{-1.02}^{+1.03}$ \\
\colrule
$\chi^2$ & \textrm{478603} & \textrm{480589} & \textrm{480789} \\
\toprule
\multicolumn{4}{c}{\textrm{Flat $\Lambda$CDM (SNe $H_0$ posterior used as $H_0$ prior)}} \\
\colrule
& $\gamma$ & $0.709_{-0.002}^{+0.002}$ &  \\
& $\beta$ & $56.828_{-0.220}^{+0.226}$ &  \\
& $H_0$ & $73.14_{-1.06}^{+1.05}$ &  \\
& $\Omega_\mathrm{m,0}$ & $0.328$ \textrm{(fixed)} &  \\
\colrule
& $\chi^2$ & \textrm{484056} &  \\
\end{tabular}
\tablecomments{$H_0$ values are given in $[\mathrm{km}\ \mathrm{s}^{-1}\ \mathrm{Mpc}^{-1}]$ units. Plots of posterior distributions are available in our public code repository~\citep{Zenodo_repo}.}
\end{ruledtabular}
\end{table}

\begin{figure}
  \includegraphics[width=\columnwidth]{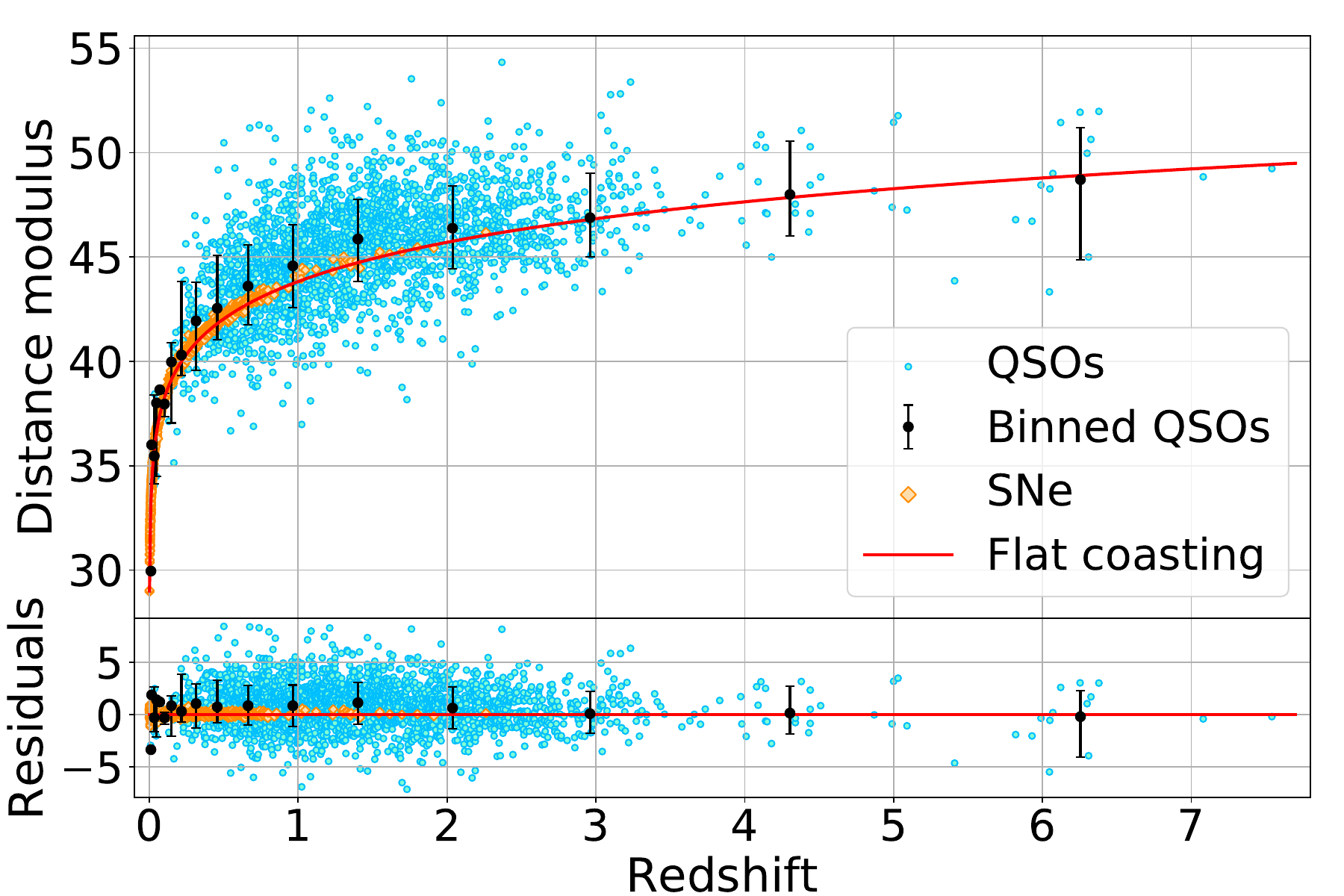}  
  \includegraphics[width=\columnwidth]{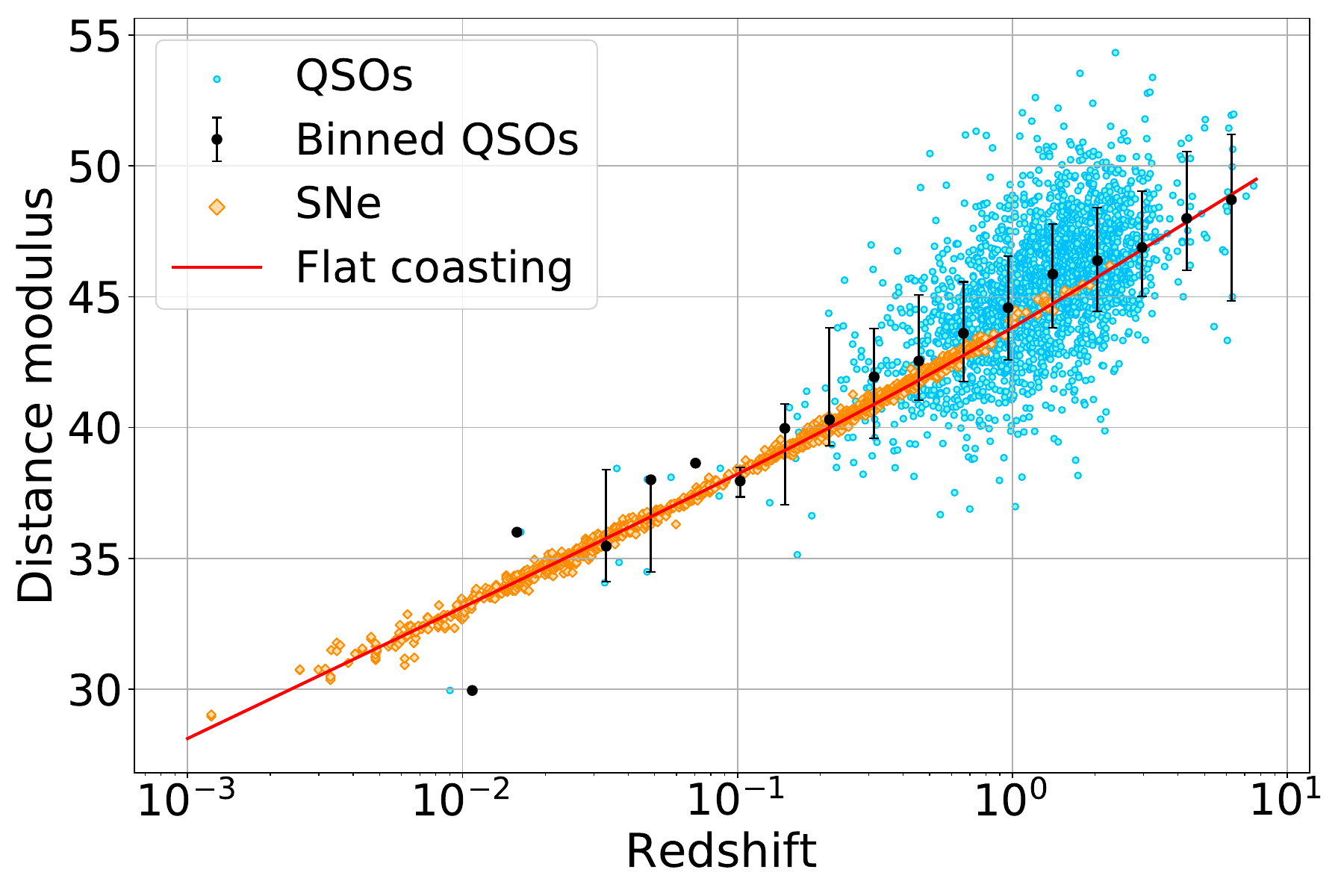}  
\caption{Distance moduli (shown without error bars) for $1701$ SN Ia observations (orange diamonds) and $2421$ QSOs (blue dots) calibrated to the flat coasting model with $H_0=71.39^{+1.02}_{-1.01}~\mathrm{km~s^{-1}~Mpc^{-1}}$ (red solid curve; see Sections~\ref{sec:SNIa} and~\ref{sec:QSO} for details). Data is from the Pantheon+ sample~\citep{Scolnic_et_al_2022} and \citet{Lusso_et_al_2020}. The normalized residuals are shown after subtracting the model $\mu(z)$ curve, along with the median $\mu_\mathrm{QSO}$ values in logarithmic redshift bins (black dots with $16$th and $84$th percentile error bars). These median $\mu_\mathrm{QSO}$ points were not used in the fitting process and are shown only for visualization. The upper and lower panels display the data on linear and logarithmic redshift scales, respectively.}
\label{fig:fig3}
\end{figure}

AD test implementations, including the one used in Sections~\ref{sec:Cosmic_chronometers} and~\ref{sec:SNIa}, are unreliable for extremely low $p$-values, such as those we obtained from the QSO data even after excluding $>4\sigma$ outliers. However, with $\Omega_\mathrm{m,0}$ fixed for the flat $\Lambda$CDM model, the fitted parameters are the same for all models, and the $H_0$ priors are also similar in width and shape. We therefore expect the differences in degrees of freedom for the four model fits to be negligible, making the $\chi^2$ values in Table~\ref{tab:table3} directly applicable for model comparison. Based on these $\chi^2$ values, the QSO data favors the coasting models over $\Lambda$CDM, with the flat coasting model being the most preferred. As a summary of the results from Sections~\ref{sec:SNIa} and~\ref{sec:QSO}, Figure~\ref{fig:fig3} shows the distance moduli for Pantheon+ SNe and \citet{Lusso_et_al_2020} QSOs, calibrated to the best-fit flat coasting model with ${H_0=71.38\pm 1.01~\mathrm{km~s^{-1}~Mpc^{-1}}}$.

\section{Conclusion}
\label{sec:Conclusion}

\begin{figure}
 	\includegraphics[width=\columnwidth]{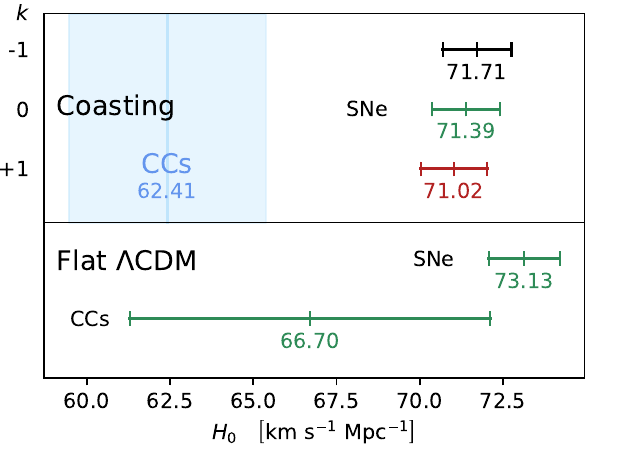}
     \caption{Best-fit $H_0$ values for coasting models with ${k=-1}$ (black), ${k=0}$ (green), and ${k=+1}$ (red) in ${H_0^2 c^{-2}}$ units (upper panel) and for the flat $\Lambda$CDM model (lower panel), based on cosmic chronometer (CCs) from \citet{Moresco_et_al_2022} and type Ia supernova (SNe) data from \citet{Scolnic_et_al_2022}. Best-fit $H_0$ values based on QSO data from \citealt{Lusso_et_al_2020} are nearly identical to the SNe results (see Table~\ref{tab:table3}) and are thus not shown. The values are the medians of the $H_0$ posteriors, with 16th and 84th percentile errors shown as error bars. There is a $\sim 3\sigma$ tension between the CC- and SN-based best-fit $H_0$ values for coasting models, while the $\Lambda$CDM model's $H_0$ values differ by $\sim 1.2\sigma$.}
     \label{fig:fig4}
\end{figure}

In Sections~\ref{sec:Cosmic_chronometers}--\ref{sec:QSO}, we presented tests and comparisons of coasting cosmological models with ${k=\left\{ -1,0,+1 \right\}}$ in ${H_0^2 c^{-2}}$ units and the flat $\Lambda$CDM model. We carried out the tests by fitting the models to $H(z)$ data derived from CCs~\citep{Moresco_et_al_2022}, and distance moduli derived from standardized SNe Ia~\citep{Scolnic_et_al_2022} and QSOs~\citep{Lusso_et_al_2020}. We used the \texttt{emcee}\footref{note:emcee_code} code~\citep{Foreman_Mackey_et_al_2013} for fitting CC data, our own MCMC implementation~\citep{Zenodo_repo} for SNe and QSOs, and AD tests for normality~\citep{Anderson_Darling_1952,adtest_2024} on normalized residuals for model comparisons. To mitigate the effects of non-cosmological contaminations, we applied sigma clipping at $3\sigma$ (e.g., \citealt{Bargiacchi_et_al_2021}) in our fitting process, and a rejection of ${>4\sigma}$ outliers in our AD tests. We carried out the SNe fits using Cepheid-calibrated host-galaxy distance moduli from \citet{Riess_et_al_2022} as external calibrators. We used the $H_0$ posteriors of the SN fits as priors for the QSO fits. Results of our fits and tests are summarized primarily in Tables~\ref{tab:table1}--\ref{tab:table3}. We also summarize the best-fit $H_0$ results for coasting models in Figure~\ref{fig:fig4}.

Coasting models are favored over the flat $\Lambda$CDM model by all datasets used, with a slight overall preference for the flat (${k=0}$) coasting model. Although none of the best-fit realizations of the four tested cosmological models, including the flat coasting model, met the $p\geq 0.05$ threshold in AD tests for consistency with the SN and QSO data, $23\%$ of flat coasting model realizations sampled from the joint posterior distribution of parameters satisfied this condition for the SN data (see Section~\ref{sec:SNIa}). The large intrinsic scatter in $\mu_\mathrm{QSO}$ data, the overfitting of the flat $\Lambda$CDM model to Pantheon+ SNe, together with the strong explanatory power of the $\Lambda$CDM model across other redshift ranges and datasets lead us to conclude that the preference for coasting models indicates a need to refine the error estimations in the SN and QSO data, rather than to revise the concordance cosmological model. At the same time, we highlight the seemingly fine-tuned nature of either the CC data or $\Omega_{\mathrm{m},0}$ in the $\Lambda$CDM model to an ${H_1=H_0}$ coincidence when fitting ${H(z)=H_1z+H_0}$ to CC data (see Section~\ref{sec:Cosmic_chronometers} for details), and recommend this for further study.

For the flat coasting model, we suggest using ${H_0=71.38\pm 1.01~\mathrm{km~s^{-1}~Mpc^{-1}}}$ (see Table~\ref{tab:table3}) as a baseline for future studies. This value is inferred from four distinct cosmological probes: CCs, SNe~Ia, QSOs, and Cepheids as SN calibrators. Note however, that this $H_0$ value shows a $2.9\sigma$ tension with ${H_0=62.41\pm 2.96~\mathrm{km~s^{-1}~Mpc^{-1}}}$ inferred from CC data alone (see Table~\ref{tab:table1}). Resolving the tension while preserving the ${H_1=H_0}$ coincidence would require the $H(z)$ values from CCs to be systematically higher (or the inferred $\Delta t$ age differences lower) by $\sim 10\%$.

The baseline $H_0$ implies an age of the universe of ${T=H_0^{-1}=13.708\pm 0.194~\mathrm{Gyr}}$ in a strictly linear flat coasting model. However, our tests allow the flat coasting model to be quasi-linear, transitioning from a $\Lambda$CDM-like expansion to coasting at redshift $z_\mathrm{t}$ between ${z^\mathrm{max}_\mathrm{QSO}\simeq 7.54}$ (the redshift of the most distant QSO in \citealt{Lusso_et_al_2020}) and ${z_*\simeq 1090}$ (the redshift at recombination; \citealt{Planck_2018}). In this scenario, with ${H_0=71.38~\mathrm{km~s^{-1}~Mpc^{-1}}}$, the present age of the universe would range from ${T_\mathrm{min}\simeq 12.27~\mathrm{Gyr}}$ for $z_\mathrm{t}=7.54$ to ${T_\mathrm{max}\simeq 13.67~\mathrm{Gyr}}$ for $z_\mathrm{t}=1090$.

\begin{acknowledgments}
The authors would like to thank Bence B\'ecsy, Istv\'an Csabai, Attila Cs\'ot\'o, Dominika E. Kis, D\'avid A. K\"odm\"on and M\'aria P\'alfi for fruitful discussions throughout the project. This project has received funding from the HUN-REN Hungarian Research Network and was also supported by the NKFIH excellence grant TKP2021-NKTA-64.
\end{acknowledgments}

\vspace{5mm}

\bibliography{Coasting_Tests}{}
\bibliographystyle{aasjournal}

\end{document}